\documentclass[a4paper,11pt]{article}
\pdfoutput=1 

\usepackage{jcappub}
\usepackage[T1]{fontenc}
\usepackage[english]{babel}
\usepackage{bm}
\usepackage{times}
\usepackage{ulem}
\usepackage{hyperref}
\hypersetup{
    colorlinks=true,
    linkcolor=blue,
    filecolor=magenta,      
    urlcolor=blue,
    citecolor=blue,
    pdftitle={Sharelatex Example},
    pdfpagemode=FullScreen
    }
\usepackage{listings}
\usepackage{slashed}
\usepackage{color}
\usepackage{appendix}
\usepackage{mathtools}
\usepackage{epsfig}
\usepackage{dcolumn}
\usepackage{textcomp}
\usepackage{appendix} 
\usepackage{multirow}
\usepackage{graphicx}
\usepackage{soul}
\usepackage{subfigure}
\newcommand{\fig}[1]{Fig.~\ref{#1}} 
 
\newcommand{\eq}[1]{Eq.~(\ref{#1})} 
\newcommand{\tab}[1]{Table~\ref{#1}}

\begin{document} 
\title{\boldmath The $H_0$ trouble: Confronting Non-thermal Dark Matter and Phantom Cosmology with the CMB, BAO, and Type Ia Supernovae data}

\author[a]{Simony Santos da Costa}
\emailAdd{simony.santosdacosta@pi.infn.it}
\affiliation[a]{Istituto Nazionale di Fisica Nucleare (INFN), Sezione di Pisa, 56127, Pisa, PI, Italy}

\author[b,1]{, D\^eivid R. da Silva~\note{Corresponding author.}}
\emailAdd{deivid.rodrigo.ds@gmail.com}
\affiliation[b]{Centro Brasileiro de Pesquisas F\'{\i}sicas, Rua Dr. Xavier Sicaud 150, Urca, CEP-22290-180, Rio de Janeiro, RJ, Brasil}

\author[c,d]{, \'Alvaro S. de Jesus}
\emailAdd{alvarosndj@gmail.com}
\affiliation[c]{Departamento de F\'isica, Universidade Federal do Rio Grande do Norte, 59078-970, Natal, RN, Brasil}
\affiliation[d]{International Institute of Physics, Universidade Federal do Rio Grande do Norte,  59078-970, Natal, RN, Brasil}

\author[b]{, Nelson Pinto-Neto}
\emailAdd{nelsonpn@cbpf.br}

\author[c,d,e]{and Farinaldo S. Queiroz}
\emailAdd{farinaldo.queiroz@ufrn.br}
\affiliation[e]{Millennium Institute for Subatomic Physics at High-Energy Frontier (SAPHIR), Fernandez Concha 700, Santiago, Chile}

\abstract{
We have witnessed different values of the Hubble constant being found in the literature in the past years. Albeit, early measurements often result in an $H_0$ much smaller than those from late-time ones, producing a statistically significant discrepancy, and giving rise to the so-called Hubble tension.  The trouble with the Hubble constant is often treated as a cosmological problem. However, the Hubble constant can be a laboratory to probe cosmology and particle physics models. In our work, we will investigate if the possibility of explaining the $H_0$ trouble using non-thermal dark matter production aided by phantom-like cosmology is consistent with the Cosmic Background Radiation (CMB) and Baryon Acoustic Oscillation (BAO) data. We performed a full Monte Carlo simulation using CMB and BAO datasets keeping the cosmological parameters $\Omega_b h^2$, $\Omega_c h^2$, $100\theta$, $\tau_{opt}$, and $w$ as priors and concluded that a non-thermal dark matter production aided by phantom-like cosmology yields at most $H_0=70.5$ km s$^{-1}$Mpc$^{-1}$ which is consistent with some late-time measurements. However, if $H_0> 72$ km s$^{-1}$ Mpc$^{-1}$ as many late-time observations indicate, an alternative solution to the Hubble trouble is needed. Lastly, we limited the fraction of relativistic dark matter at the matter-radiation equality to be at most 1\%. }


\maketitle
\flushbottom

\section{\label{In} Introduction}

The value of the Hubble parameter is highly dependent on the techniques used to measure it. Typically, early measurements of this parameter yield values that are statistically inconsistent with late time measurements, giving rise to the so-called Hubble tension \cite{DiValentino2021}. Early measurements involve techniques that extract information from the very early universe, whereas late measurements rely on data from astrophysical objects, such as stars, at a phase where cosmic structures have already formed.

To grasp an understanding of the tension, we briefly review some important late-time measurements. SHOES Collaboration used Cepheids to calibrate Type Ia supernovae luminosity, resulting in $H_0 = 73.2 \pm 1.3$ km s$^{-1}$ Mpc$^{-1}$ with an error of $1.8\%$ \cite{Riess:2020fzl}. The Carnegie–Chicago Hubble Programme (CCHP) combined data from red giants and Type Ia supernovae to obtain $H_0 = 69.8 \pm 0.6 \text{(stat)} \pm 1.6 \text{(sys)}$ km s$^{-1}$ Mpc$^{-1}$ \cite{Freedman:2021ahq}. HOLiCOW Collaboration employed time-delay techniques in six gravitationally lenses, which produced $H_0 = 73.3 \pm 1.7$ km s$^{-1}$ Mpc$^{-1}$ \cite{Wong:2019kwg}. Gravitational-wave events resulting from the merger of neutron stars were detected by LIGO and Virgo, providing an estimate of $H_0 = 70^{+12}_{-8}$ km s$^{-1}$ Mpc$^{-1}$ \cite{LIGOScientific:2017adf}. Additional standard sirens from these events were included in the analysis, yielding $H_0 = 72^{+12}_{-8}$ km s$^{-1}$ Mpc$^{-1}$ \cite{DES:2020nay}. Parallax measurements of Cepheids gave rise to $H_0 = 73.24 \pm 1.74$ km s$^{-1}$ Mpc$^{-1}$ \cite{ref:Bernal2016}. The overlap of Cepheids and Type Ia supernovae (SNIa) provides a range for $H_0$ between $70.1$ and $77.1$ km s$^{-1}$ Mpc$^{-1}$ \cite{DiValentino2021}. Combining the tip of the red giant branch (TRGB) with SNIa measurements yields a range for $H_0$ between $67.7$ and $74.2$ km s$^{-1}$ Mpc$^{-1}$ \cite{DiValentino2021}. The Tully-Fisher relation (TFR), which correlates the luminosity of a galaxy with its rotational velocity, has also been used to estimate $H_0$, resulting in a range of $72.3$ to $78.6$ km s$^{-1}$ \cite{DiValentino2021}. Hence, it is clear that several late-time measurements prefer a value of $H_0> 72$ km s$^{-1}$ Mpc$^{-1}$, with few reporting a value of $H_0$  around  $72$ km s$^{-1}$ Mpc$^{-1}$.

As for early-time measurements of the Hubble constant, they come essentially from the power spectrum of the CMB. Planck data alone favors $H_0 = 67.27 \pm 0.6$ km s$^{-1}$ Mpc$^{-1}$ \cite{ref:Planck2018, DiValentino2021}, which is statistically discrepant from the value extracted from late-time observations. Therefore, early-time measurements of the Hubble constant favor $H_0 < 69$ km s$^{-1}$ Mpc$^{-1}$, whereas local measurements prefer $H_0 > 70$ km s$^{-1}$ Mpc$^{-1}$ \cite{ref:Anchordoqui2021, DiValentino2021}. The discrepancy in the values of the Hubble constant is not the core issue; rather, it is the statistical incompatibility that poses the real problem. There are late and early measurements that differ by more than $5\sigma$ from each other \cite{DiValentino2021,ref:Verde2019}.

Consequently, either early-time measurements need to be revised or the methods astronomers use to calculate distances of very distant objects need to be improved. Though direct, the late-time measurements are much more subject to systematic errors than the early-time ones. As mentioned above, when using the method called Tip of the Red Giant Branch (TRGB) \cite{Freedman:2021ahq}, the value inferred for $H_0$ ($H_0 = 69.8 \pm 0.6 \text{(stat)} \pm 1.6 \text{(sys)}$ km s$^{-1}$ Mpc$^{-1}$) is bigger but compatible with the early universe evaluations, within the error bars. In many cases, the red giants and Cepheids used in these datasets belong to the same galaxies, indicating the presence of large systematic errors perhaps connected with the physics of the Cepheids, as their metallicity properties, see a discussion of this conflict in Ref.~\cite{Freedman:2021ahq}. Ultimately, the new data from the James Webb Space Telescope, as well as a relatively new method, the J-Region Asymptotic Giant Branch method, may either eliminate this $H_0$ tension, see Ref.~\cite{Freedman:2023ahq} for details. Alternatively, the $H_0$ problem represents an augury for new physics. 

We assume the Hubble tension has a physical nature in the early universe.  We tackle it through a new physics episode in the early universe involving dark matter particles. In this way, early-time measurement of the Hubble constant is influenced by it, leading to a higher value of $H_0$, in the direction of values extracted from late-time observations \cite{ref:Anchordoqui2021,Kenworthy:2022jdh}. In particular, we consider a non-thermal production mechanism of dark matter particles that produces a boost factor larger than one, thus increasing the relativistic energy density in the early universe, which in turn has the potential to raise $H_0$ and ultimately alleviate the tension \cite{ref:Hooper2011,ref:Kelso2013, Deivid:2022,Deivid:2023}. In other words, this setup increases the number of relativistic degrees of freedom, i.e. $\Delta N_{eff} > 0$, and then raises $H_0$ at early-times  \cite{ref:Planck2018, ref:Anchordoqui2021, ref:Bernal2016, SunnyVagnozzi2020}.

The mechanism relies on the assumption that an unstable and heavy particle $\chi^\prime$ thermally decoupled from the fundamental plasma in the early universe, and then eventually decayed into a dark matter $\chi$ particle plus a standard particle, a neutrino or a photon, for instance. If $\chi^\prime$ it is much heavier than $\chi$, then $\chi$ will be relativistic, in contradiction with all cosmological data that suggest a prominent cold dark matter component. In fact, if a large fraction of the overall dark matter abundance comes from the decay of $\chi^\prime$, the matter power spectrum changes significantly, leading to disagreements with Lyman-$\alpha$ observations \cite{Allahverdi:2014bva}. This small abundance consideration is essential to avoid conflicts with structure formation \cite{Bringmann:2018jpr,Feng:2003uy, ref:Cyburt2003}. Indeed, using data from the Sloan Digital Sky Survey galaxy cluster, it has been found that at most 6\% of the overall dark matter abundance can be relativistic when structures start to be formed \cite{Reid:2009xm,Zhao:2012xw,Allahverdi:2014bva}. Thus, we circumvent this issue, assuming that only a small fraction of the dark matter abundance arises from this decay.  Indeed, the dark matter particles produced by the decay are relativistic for a period, but as the universe expands, they cool and become a standard ensemble of cold particles. Assuming this fraction to be of relativistic dark matter particle to be small, we obey the constraints from structure formation. Additionally, having a small fraction of dark matter particles being relativistic at the matter-radiation equality induces a suppression in the matter power spectrum due to a larger free-streaming length, which is needed to solve the missing satellite problem as presented in the context of sterile neutrinos and other new physics scenarios \cite{Cembranos:2005us,Strigari:2006jf,Strigari:2007ma,Tollerud:2008ze}.

Although, it has already been demonstrated that the introduction of $N_{\text{eff}}$ as an extra parameter into CMB analyses is not sufficient to increase $H_0$ to a point where the Hubble tension is fully solved \cite{ref:Anchordoqui2021, SunnyVagnozzi2020}. Hence, besides this particle physics process, we will invoke a phantom-like cosmology \cite{ref:Anchordoqui2021, SunnyVagnozzi2020, Abdalla:2022yfr}.

We must point out that phantom dark energy is theoretically problematic. Taking scalar fields as the phantom field, either its Hamiltonian must have a negative kinetic term, which leads to vacuum instabilities as the Hamiltonian is not bounded from below, or, without a negative kinetic term, the theory is not unitary. However, theoretical models have been constructed in order to circumvent this problem \cite{Matsumoto2018, Ludwick2018}. Regarding the adoption of phantom-like fluid, we treat it in a phenomenological manner, checking that it is consistent with the data and does not conflict with well-established physical measurements. 

In summary, we test if this phantom-like cosmology plus a non-thermal dark matter production suffices to alleviate the Hubble tension using CMB and BAO datasets. Our work is structured as follows: In Section \ref{sec:sectionII}, we connect the dark matter production with $N_{eff}$; in Section \ref{CMB}, we present our results from Monte Carlo simulations before drawing our conclusions in Section \ref{dc}; in Appendix \ref{sec:entropy_injection}, we assess the entropy injection of the mechanism.


\section{Increase in relativistic energy density produced by dark matter}\label{sec:sectionII}

Suppose that after neutrino decoupling but before matter-radiation equality, a heavy particle $\chi^\prime$ decays into dark matter $\chi$ and massless neutrinos $\nu$, with $m_{\chi^\prime} \gg m_{\chi}$. We are also assuming that neutrinos are massless particles. The produced dark matter is boosted, and as aforementioned we assume that only a small fraction of the dark matter abundance is produced through this mechanism to avoid conflict with structure formation data. Initially, the dark matter is in a hot state, contributing relativistic energy density. Over time, it cools down and becomes cold. The extra neutrinos have decoupled from the primordial plasma and behave like a perfect relativistic fluid, with its energy density decaying inversely with the fourth power of the scale factor, i.e., $\rho_{\mathrm{extra}} \propto 1/a^4$ \cite{dodelson2020}. Therefore, the current energy density is the sum of background photons and neutrinos \cite{dodelson2020, ref:Planck2018}, along with the additional relativistic energy density introduced by the decay of $\chi^\prime$,
\begin{equation}
    \rho_{\mathrm{r, 0}} = \rho_{\gamma, 0} + N_\nu \times \rho_{1\nu, 0} + \rho_{\mathrm{extra, 0}},
\end{equation}where $N_\nu$ represents the number of relativistic species in the SM, which is basically the number of neutrino species, and $\rho_{1\nu}$ is the energy density of a single neutrino species.

Therefore, it is useful to parameterize the extra radiation source as,
\begin{equation}
    \rho_{\text{extra}, 0} = \Delta N_{\text{eff}} \times \rho_{1\nu, 0}.
\end{equation}
This allows us to express the total radiation energy density today as,
\begin{equation}
    \rho_{\mathrm{r, 0}} = \rho_{\gamma, 0} + N_{\text{eff}} \times \rho_{1\nu, 0}.
\end{equation}
In this equation, the term $N_{\text{eff}} = 3 + \Delta N_{\text{eff}}$ serves as an additional parameter to be determined through likelihood analyses of the CMB power spectrum. Taking into account the time evolution of radiation energy allows us to rewrite $\Delta N_{\mathrm{eff}}$ as,
\begin{equation}
   \Delta  N_{\mathrm{eff}} = \frac{\rho_{\text{extra}, 0} / a^4_{\mathrm{eq}}}{\rho_{1\nu, 0} / a^4_{\mathrm{eq}}} = \frac{\rho_{\text{extra}}(t_\mathrm{eq})}{\rho_{1\nu}(t_\mathrm{eq})} \cdot
   \label{eq:deltaN_general}
\end{equation}

We point out that a $\Delta  N_{\mathrm{eff}}\neq 0$  has been found in recent analyses of the CMB data \cite{ref:Anchordoqui2021} supporting the idea of an extra radiation energy component. In our case, this extra radiation counterpart stems from the $\chi^\prime$ decay \cite{ref:Hooper2011, ref:Kelso2013, ref:Alcaniz2021}.

To calculate $\Delta N_{\mathrm{eff}}$ rising from the  $\chi^\prime$ decay, we need to consider a concrete decaying setup. We will assume that the decay is of the type $\chi^\prime \rightarrow \chi + \nu$. In the $\chi^\prime$ rest frame, the 4-momenta of the particles involved in the decay are as follows,
\begin{gather*}
    p_{\chi^\prime} = (m_{\chi^\prime}, \bm{0}),\\
    p_{\chi} = (E_{\chi}(\bm{p}), \bm{p}),\\
    p_{\nu} = (|\bm{p}|, -\bm{p}).
\end{gather*}

Therefore, the 4-momentum conservation gives the energy and momentum at the moment immediately after decay,
\begin{gather}
    \left|\bm{p}_{\chi}(\tau) \right| = \left|\bm{p} \right| = \frac{1}{2} m_{\chi^\prime} \left[ 1 - \left(\frac{m_{\chi}}{m_{\chi^\prime}} \right)^2 \right], \label{eq:momentum_chi}\\
    E_{\chi}(\tau) = m_{\chi} \left( \frac{m_{\chi^\prime}}{2m_{\chi}} + \frac{m_{\chi}}{2m_{\chi^\prime}} \right), \label{eq:energy_in_tau}
\end{gather}
where $\tau$ is the lifetime of $\chi^\prime$.

Setting the Lorentz factor to be,
\begin{equation}
    \gamma_{\chi}(\tau) = \left( \frac{m_{\chi^\prime}}{2m_{\chi}} + \frac{m_{\chi}}{2m_{\chi^\prime}} \right),
\end{equation}we can rewrite \eq{eq:energy_in_tau} and find,
\begin{equation}
    E_{\chi}(\tau) = m_{\chi}\gamma_{\chi}(\tau).
\end{equation}

After the $\chi^\prime$ decay, the $\chi$ momentum obeys the relation $\bm{p}^2_{\chi} \propto 1/a^2$ \cite{ref:hobson2006GR}, which implies in,
\begin{equation*}
    \begin{split}
        &E^2_{\chi} - m^2_{\chi} = \bm{p}^2_{\chi} \propto \frac{1}{a^2}\\
        &\Rightarrow  \left( E^2_{\chi}(t) - m^2_{\chi} \right)a^2(t) = \left( E^2_{\chi}(\tau) - m^2_{\chi} \right)a^2(\tau)\\
        &\Rightarrow E_{\chi}(t) = m_{\chi}\left[ 1 + \left( \frac{a(\tau)}{a(t)} \right)^2 \left( \gamma^2_{\chi}(\tau) - 1\right)\right]^{1/2}.
    \end{split}
\end{equation*}

We are considering a radiation-dominated phase, in which $a \propto \sqrt{t}$ \cite{ref:hobson2006GR}, thus $a(\tau)/a(t) = \sqrt{\tau/t}$. Therefore, the Lorentz boost factor at time $t$ is
\begin{equation}
    \gamma_{\chi}(t) = \sqrt{\frac{ (m^2_{\chi} - m^2_{\chi'})^2 }{4m^2_{\chi}m^2_{\chi'}} \left( \frac{\tau}{t} \right) + 1 }.
    \label{eqgamma}
\end{equation}

With Eq. (\ref{eqgamma}) we find the energy of dark matter particle to be,
\begin{equation}
    E_{\chi} = m_{\chi}(\gamma_{\chi} -1) + m_{\chi}.
\end{equation}
When the second term on the right-hand side of this equation dominates, the dark matter particle is non-relativistic. The first term is then interpreted as the mean contribution to the particle energy in the ultrarelativistic regime.

We are considering that our formalism generates only a small fraction of the dark matter particles. Hence, the total energy of the dark matter particles is,
\begin{equation}
    E_{\mathrm{DM}} = N_{\mathrm{HDM}}m_{\chi}(\gamma_{\chi} -1) + N_{\mathrm{CDM}}m_{\chi},
\end{equation}
where $N_{\mathrm{HDM}}$ is the number of hot dark matter particles, $N_{\mathrm{CDM}}$ is the number of cold dark matter particles, and they obey the relation $N_{\mathrm{HDM}} \ll N_{\mathrm{CDM}}$.

The ratio between the two dark matter energy densities is,
\begin{equation}
    \frac{\rho_{\mathrm{HDM}}}{\rho_{\mathrm{CDM}}} = \frac{n_{\mathrm{HDM}}m_{\chi}\left( \gamma_{\chi} -1 \right)}{n_{\mathrm{CDM}}m_{\chi}} \equiv f\left( \gamma_{\chi} -1 \right),
    \label{Eq.ratio}
\end{equation}
where $n_{\mathrm{HDM}}$ and $n_{\mathrm{CDM}}$ represent the number densities of relativistic and nonrelativistic dark matter particles, respectively. The factor $f$ denotes the ratio between these two number densities, and it must be small. 

\begin{figure*}[htb!]
    \centering
    \includegraphics[width = 0.49\linewidth]{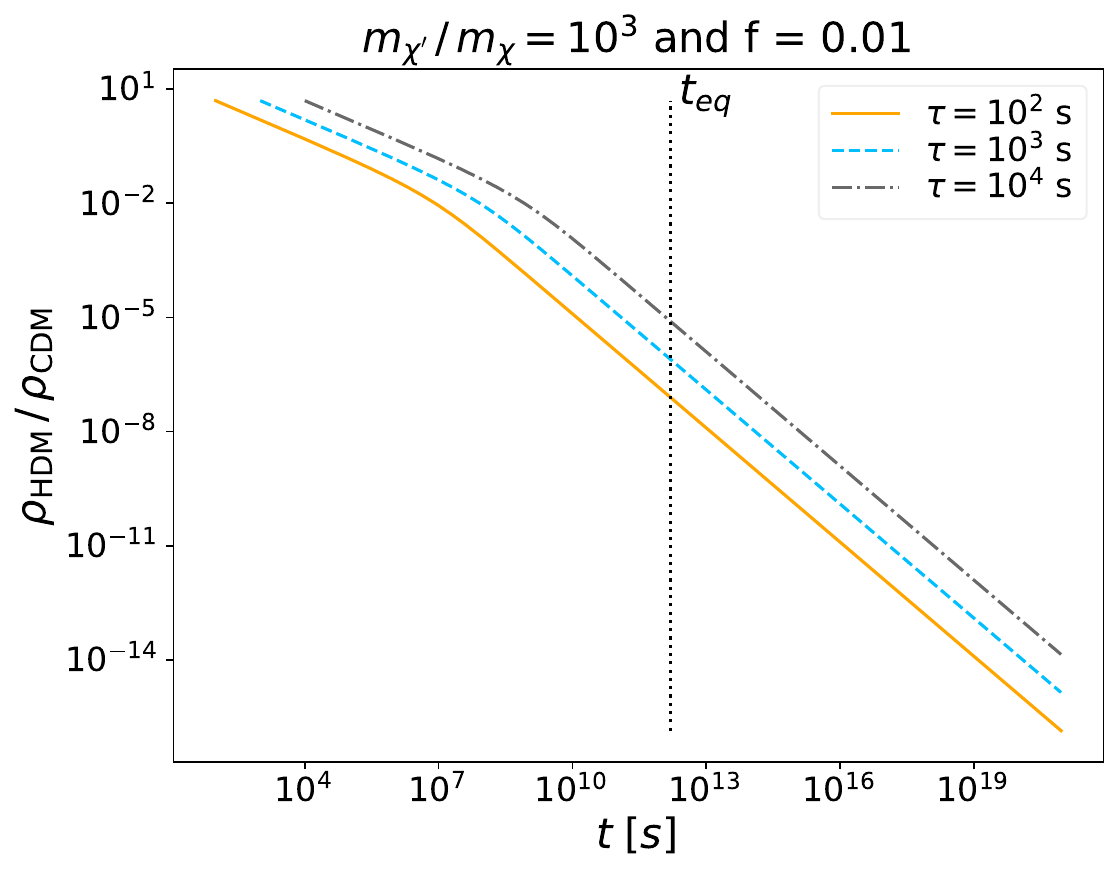}
    \includegraphics[width = 0.49\linewidth]{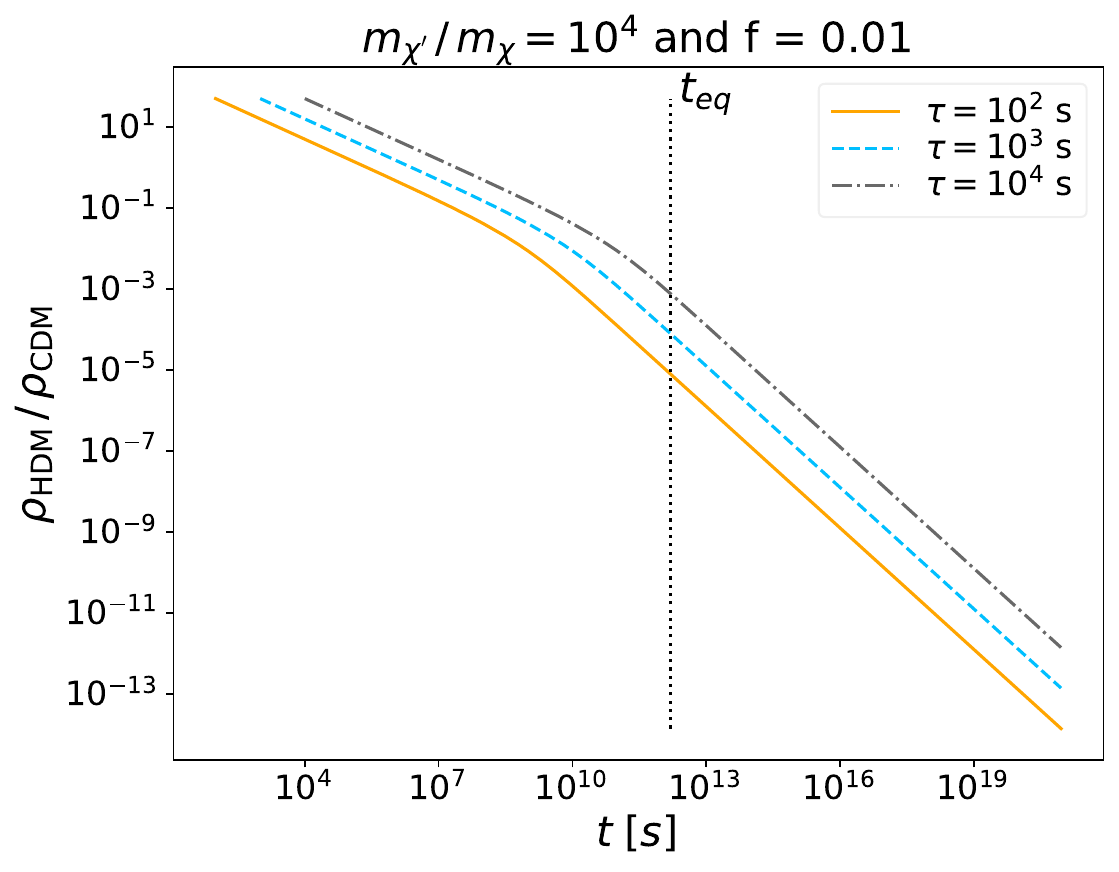}
    \includegraphics[width = 0.49\linewidth]{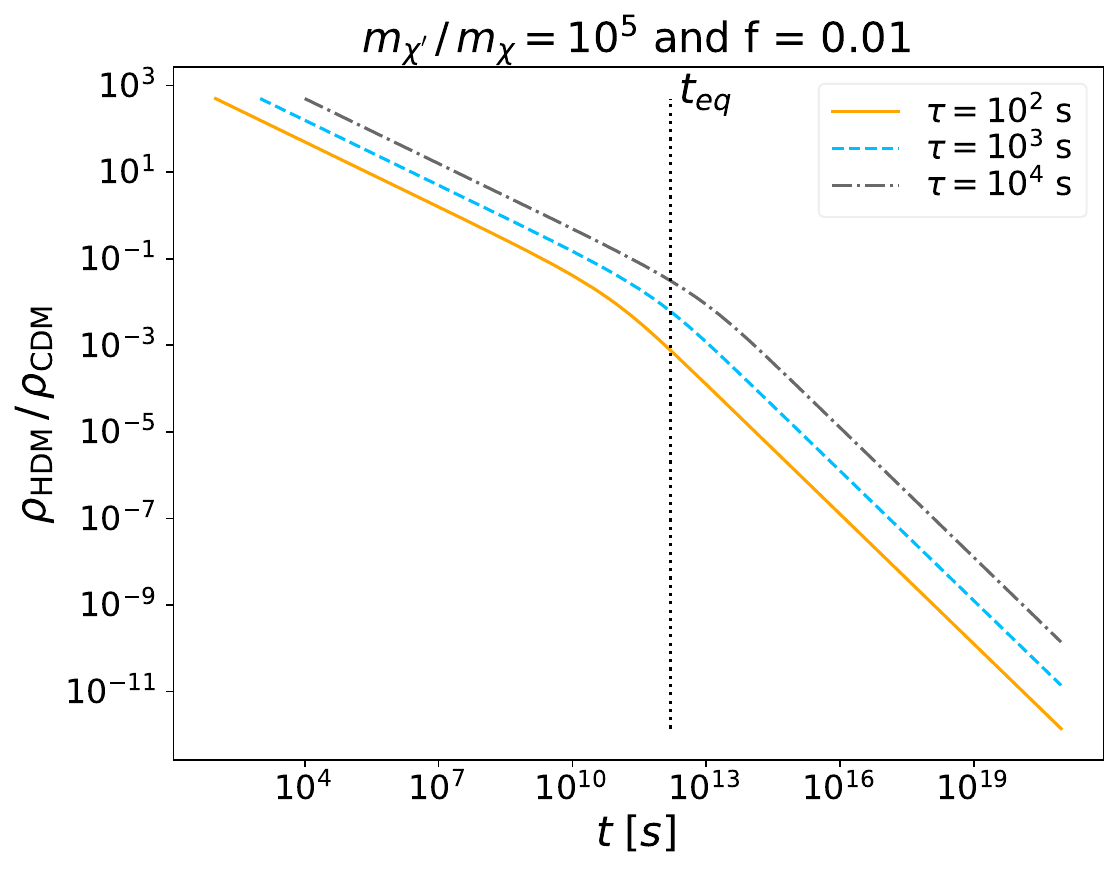}
    \includegraphics[width = 0.49\linewidth]{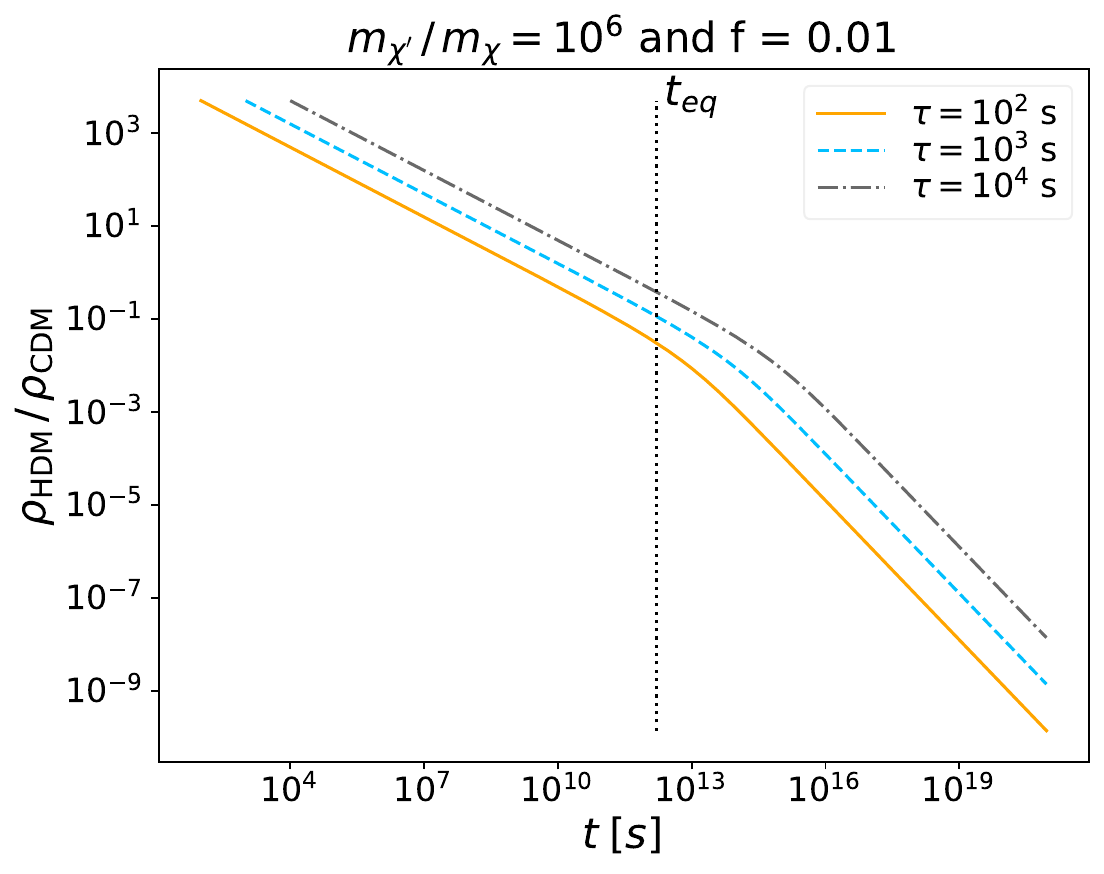}
    \caption{The time evolution of the hot dark matter and cold dark matter ratio is shown in the plots. The yellow solid line represents the case where the lifetime of $\chi^\prime$ is $\tau = 10^2$ s, while the blue dashed line represents the case where $\tau = 10^3$ s, and the gray dotted line represents the case where $\tau = 10^4$ s. In all cases, it was assumed that only 1\% of the dark matter is formed by $\chi^\prime$ decay, i.e., $f = 0.01$. This f value is used for illustration purposes. We also vary the mass ratio between the dark matter mother and the dark matter itself to illustrate how the energy ratio changes with it.}
    \label{fig:t_x_rho_ratio}
\end{figure*}

In Fig. \ref{fig:t_x_rho_ratio}, we illustrate how this hot-cold energy density ratio changes over time. In this figure, we consider the illustrative case of $f = 0.01$ and vary the mass ratio $m_{\chi^\prime} / m_{\chi}$ in the range of $10^3$ to $10^6$. We also vary the lifetime of $\chi^\prime$ between $10^2$ and $10^4$ seconds. This range for the lifetime was chosen to avoid conflicts with the Big Bang Nucleosynthesis data. Longer lifetimes are excluded because the injection of relativistic particles at late times alters the abundance of the light elements \cite{Deivid:2022}. These values for the mass ratio and lifetime are the prior values used in the CMB analyses in Sec. \ref{CMB}. Fig. \ref{fig:t_x_rho_ratio} also shows that the larger the ratio $m_{\chi^\prime} / m_{\chi}$, the greater the hot dark matter energy density as time passes, in agreement with Eq. (\ref{eq:energy_in_tau}) and Eq. (\ref{Eq.ratio}). Moreover, the longer the lifetime the larger the dark radiation at matter-radiation equivalence. 

Assuming that the extra source of radiation in \eq{eq:deltaN_general} comes from boosted dark matter particles we get,
\begin{equation}
    \Delta N_{\mathrm{eff}} = \lim_{t \to t_{\mathrm{eq}}} \frac{\rho_{\mathrm{HDM}}}{\rho_{1\nu}} = \lim_{t \to t_{\mathrm{eq}}} \frac{\rho_{\mathrm{CDM}}f(\gamma_{\chi} - 1)}{\rho_{1\nu}}\cdot
\end{equation}

The ratio between the energy density of cold dark matter and that of one neutrino at the matter-radiation equality is,
\begin{equation}
    \left. \frac{\rho_{1\nu}}{\rho_{CDM}} \right|_{t = t_{eq}} = \frac{\Omega_{\nu,0}\rho_c}{3a^4_{eq}} \times \left( \frac{\Omega_{CDM,0}}{a^3_{eq}}  \right)^{-1} = 0.16,
    \label{eq:0.16}
\end{equation}where we used $\Omega_{\nu,0} = 3.65 \times 10^{-5}$, $\Omega_{\mathrm{CDM},0} = 0.265$, and $a_{\mathrm{eq}} = 3 \times 10^{-4}$ \cite{Aghanim:2018eyx}. Consequently, 
\begin{equation}
    \Delta N_{eff} = \lim_{t \to t_{eq}} \frac{f\left( \gamma_{\chi} -1 \right)}{0.16}\cdot
\end{equation}
We can find a more simplified version of this expression, taking advantage of that $m_{\chi^\prime} \gg m_{\chi}$. In this regime, 
\begin{equation}
    \gamma_{\chi}(t_{eq}) -1 \approx \gamma_{\chi}(t_{eq}) \approx \frac{m_{\chi^\prime}}{2m_{\chi}} \sqrt{\frac{\tau}{t_{eq}}},
\end{equation}leading to,
\begin{equation}
        \Delta N_{eff} \approx 2.5 \times 10^{-3}\sqrt{\frac{\tau}{10^{6} \mathrm{s}}} \times f\frac{m_{\chi'}}{m_{\chi}},
    \label{eq:deltaN}
\end{equation}where we used the following information: $t_{\mathrm{eq}} \approx 50,000$ years \cite{tongLecturesCosmology}, which is approximately $1.6 \times 10^{12}$ seconds.

It is clear that the cosmological parameter $\Delta N_{\mathrm{eff}}$ depends solely on the particle physics properties of $\chi^\prime$ and $\chi$. $\Delta N_{eff}$ grows with the lifetime of the mother particle and the mass ratio. Notice that our mechanism induces $\Delta N_{\mathrm{eff}}\neq 0$ without invoking a fourth-generation neutrino. As this non-thermal production involves highly boosted particles, we show in the Appendix \ref{sec:entropy_injection} that it does not lead to a large entropy injection in the universe which avoids problems with Big Bang nucleosynthesis.
%

Interestingly, there is a positive correlation between $\Delta N_{\mathrm{eff}}$ and the Hubble parameter ($H_0$) \cite{ref:Bernal2016}. In other words, the formula mentioned above could potentially increase $H_0$ \cite{ref:Alcaniz2021, ref:Hooper2011, Deivid:2022, Deivid:2023}. However, in our work, we will assess whether this extra source of radiation can indeed raise $H_0$ and alleviate the Hubble tension without altering much other well-known cosmological parameters. To do so, we will perform a full Monte Carlo simulation, as we describe in the next section. 

\section{\label{CMB}Observational constraints}

We have thus far addressed how to produce $\Delta N_{eff}\neq 0$ via the non-thermal production of dark matter particles. We have included $\Delta N_{eff}$ from Eq.\eqref{eq:deltaN} into the CAMB code~\cite{Lewis:1999bs}, which is a cosmological code for computing the power spectrum of Cosmic Microwave Background temperature fluctuations, by solving the Boltzmann equations and evolving the background expansion of all species considered in the model. As we are changing the composition of the relativistic species energy density, we will change the early expansion history and, consequently, the $H_0$ value. We realized that this mechanism alone cannot alleviate the Hubble tension without altering much other well-known cosmological parameters. Therefore, we realized that this dark matter production mechanism must be aided by a phantom-like model in order to alleviate the Hubble tension. Thus, we carried out our simulation using a phantom fluid plus non-thermal dark matter model.

At this stage, we linked our modified version of CAMB with the CosmoMC code~\cite{Lewis:2002ah}, a Markov Chain Monte Carlo code necessary to explore the full cosmological parameter space. We performed two different analyses: (i) work with the combination $f\frac{m_{\chi'}}{m_{\chi}}$ as a unique free parameter, and (ii) kept only $f$ free to float and fixed the ratio $\frac{m_{\chi'}}{m_{\chi}}$. In both cases, we fixed $\chi'$ mass and ran the code for different lifetime values $\tau=10^2$s, $\tau=10^3$s, and $\tau=10^4$s\footnote{Those benchmark points were selected to avoid inconsistencies with BBN, CMB power spectrum, and structure formation~\cite{ref:Alcaniz2021,ref:Hooper2011,ref:Kelso2013}.}. We also vary the usual cosmological variables, namely the baryon and the cold dark matter density, the ratio between the sound horizon and the angular diameter distance at decoupling, the optical depth and the equation-of-state parameter: $\{\Omega_b h^2, \Omega_c h^2, 100\theta, \tau_{opt}, \omega\}$. Further, we consider purely adiabatic initial conditions, fix the sum of neutrino masses to $0.06$ eV and the universe curvature to zero, and also vary the nuisance foreground parameters~\cite{Planck:2019nip}. The flat priors on the cosmological parameters used in our analysis are shown in Table \eqref{tab:priors}.

\begin{table}
\centering
\caption{Priors on the cosmological parameters considered in the analysis.}
{\begin{tabular}{|c|c|}
\hline 
Parameter & Prior Ranges \\ 
\hline
$\Omega_{b}h^{2}$ & $[0.005 : 0.1]$ \\ 
 
$\Omega_{c}h^{2}$ & $[0.001 : 0.99]$ \\ 

$100\theta$ & $[0.5 : 10.0]$ \\ 
 
$\tau_{opt}$ & $[0.01 : 0.8]$ \\ 
  
$\omega$ & $[-3 : -1]$ \\ 

$\log_{10}\left(f\frac{m_{\chi'}}{m_{\chi}}\right)$ & $[ 
-4 : 2]$ \\

$\log_{10}f$ & $[-6 : -1]$\\
\hline 
\end{tabular}\label{tab:priors}}
\end{table} 

We compared our theoretical predictions with the most recent CMB temperature data (Planck 2018) combined with observations of Baryonic Acoustic Oscillations (BAO) from the 6dF Galaxy Survey (6dFGS)~\cite{Beutler:2011hx}, Sloan Digital Sky Survey (SDSS) DR7 Main Galaxy Sample galaxies~\cite{10.1093/mnras/stv154}, BOSS galaxy samples, LOWZ and CMASS~\cite{Zhao:2015gua}, and type Ia Supernovae (SNeIa) from Pantheon collaboration~\cite{Pan-STARRS1:2017jku}.

\begin{figure*}[tbp]
    \centering
    \subfigure[]{
    \includegraphics[width = .47\linewidth]{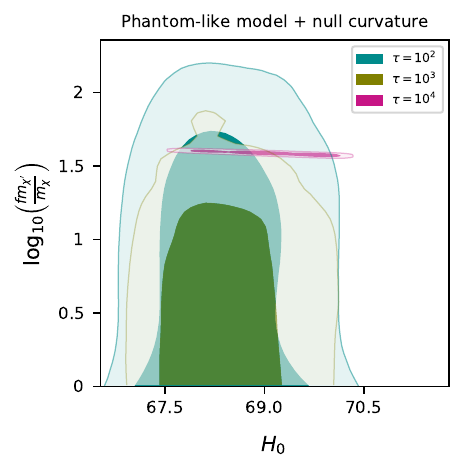}
    \label{fig:fmmxH0}
    }
    \subfigure[]{
    \includegraphics[width =.47\linewidth]{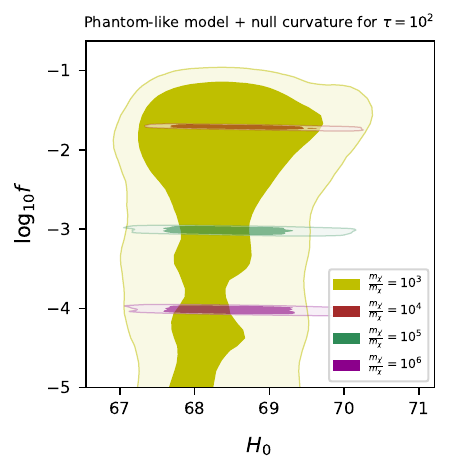}
    \label{fig:fxH0_tau2}
    }
    \subfigure[]{
    \includegraphics[width =.47\linewidth]{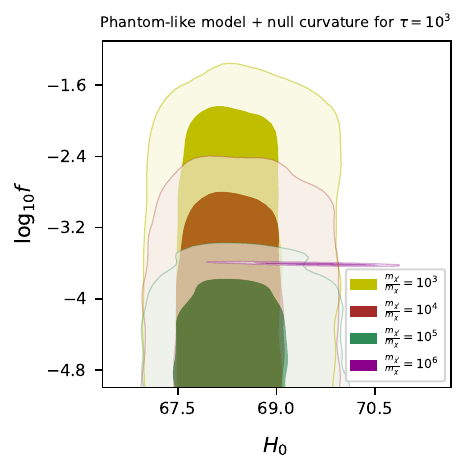}
    \label{fig:fxH0_tau3}    
    }
    \subfigure[]{
    \includegraphics[width =.47\linewidth]{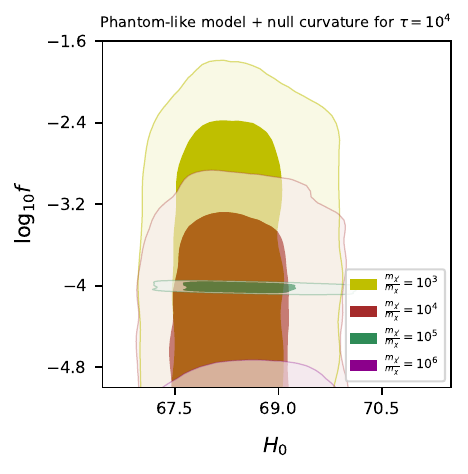}
    \label{fig:fxH0_tau4}    
    }
     \caption{Allowed regions of parameters that connect the decay mechanism and the value of Hubble constant in phantom-like cases. \textbf{(a)} The contour regions correspond to cases where $\chi^\prime$ lifetime is $10^2$s, $10^3$s, or $10^4$s, with $95 \%$ of C.L. (the huge and lighter regions) and 
     $68 \%$ of C.L. (the small and darker regions). It is a cosmology with no curvature, where exists a phantom-like quintessence and non-zero $\Delta N_{eff}$. The bounds use Planck 2018 CMB data, BAO, and type Ia data from the Pantheon sample. \textbf{(b)} This case is also a universe with phantom-like quintessence and $\Delta N_{eff}$, but considering a fixed value for lifetime decay of $\tau=10^2$ and different values for the mass ratio $\frac{m_{\chi'}}{m_{\chi}}$. \textbf{(c)} and \textbf{(d)} are the same as (b) but consider a fixed value for the lifetime decay of $\tau=10^3$ and $\tau=10^4$, respectively. The $H_0$ parameter is measured in [km s$^{-1}$ Mpc$^{-1}$] in all plots of this work.}
    \label{fig:fxH0}
\end{figure*}

\begin{figure*}[tbp]
\centering
\includegraphics[width=\textwidth]{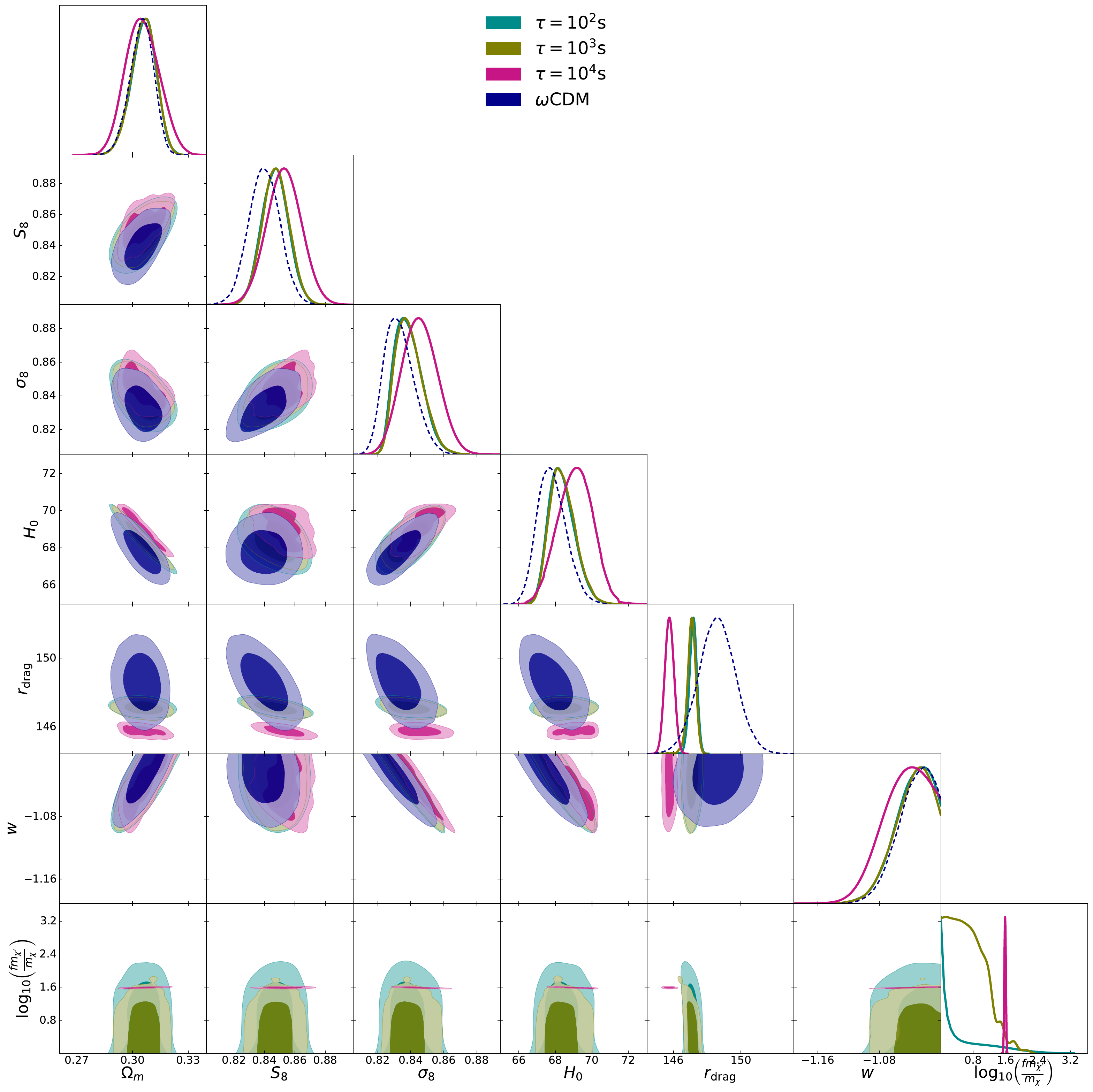}
\caption{One-dimensional posterior distributions and two-dimensional joint contours for the parameter space $\{\Omega_bh^2, \Omega_ch^2, 100\theta, \tau$, $\omega, f\frac{m_{\chi'}}{m_{\chi}}\}$ for Planck 2018 + BAO + Pantheon dataset. The $r_{drag}$ parameter is measured in [Mpc] in all the triangular plots of this work. We display also the $\omega$CDM model for comparison purposes.}
\label{fig:tri_tau234_fmm}
\end{figure*}

In ~\fig{fig:fmmxH0}, we show the $H_0 \times \left(fm_{\chi^\prime} / m_{\chi}\right)$ plane, using Phantom-like models. We have assessed the impact of having a non-zero curvature. However, no improvements were found when we considered the curvature free to float. Hence, we chose to display only the standard case of null curvature. Note that the best case is the one for $\tau=10^4 s$, where we can severely constrain the combination $fm_{\chi^\prime} / m_{\chi}$ with excellent error bars. This is due to the upper limit of $\tau$, which is the maximum value still respecting BBN limits \cite{Deivid:2022}, in such a way that we remain with a smaller allowed region in agreement with the data. The complete list of cosmological parameters for the analysis considering the combination $\left(f\frac{m_{\chi'}}{m_{\chi}}\right)$ free is shown in Tab.\eqref{tab:Tabel_results_1}, where we considered the jointed dataset of Planck 2018+BAO+SNeIa. We also presented the constraints for the $\omega$CDM + $N_{eff}$ model for comparison purposes in the first column block. It is important to mention the full agreement with $\omega$CDM$+N_{eff}$ model in $1\sigma$ C.L., with two main advantages: first, we explain the origin for the change in $N_{eff}$ due to the decay mechanism for producing dark matter, and second by increasing the $H_0$ value to $69.08\pm 0.71$ we reduce to $3.2\sigma$ the tension with local measurements (see third column of Tab \eqref{tab:Tabel_results_1}. For completeness, we show in \fig{fig:tri_tau234_fmm} the contour regions for the most interesting parameters of the non-thermal DM models compared with the $\omega$CDM + $N_{eff}$ model, where we can check the agreement in $1\sigma$ C.L. \footnote{The $\tau$ presented in the parameter space of \fig{fig:tri_tau234_fmm} corresponds to the optical depth and it is different from the lifetime $\tau$ present in the legend of all plots in this work.}. An exception occurs for $\tau=10^4 s$, where we have an agreement only in $2\sigma$ for $r_{drag}$\footnote{The baryonic acoustic scale value calculated using the sound horizon at the time the baryon velocity decouples from the photons.}. Indeed, the smaller value for the acoustic scale $r_{drag}$ is justified with the increase on $H_0$, since they are negatively correlated. Fortunately, this anti-correlation is not enough to alter substantially the $\sigma_8$\footnote{The amplitude of matter density fluctuations when considering spheres of radius $R=8$Mpc$/H_0/100$.}, and therefore we do not increase the tension on the clustering parameter.


\begin{table*}
\centering
\caption{$68\%$ confidence limits for the cosmological parameters. The first column block shows the constraints for the $\omega$CDM model, while the second one refers to the non-thermal DM model, for each dark matter mother particle lifetime considered, $\tau_{life}$. The constraints on the parameters of the model used the extended data set, i.e. the joint Planck 2018 + BAO + Pantheon dataset. 
Note that the table is divided into two sections: the upper section shows the primary parameters, while the lower part shows the derived ones.}
\scalebox{0.9}{
{\begin{tabular}{|c|c|c|c|c|}
\hline
 & {$\omega$CDM + $N_{eff}$}
 &\multicolumn{3}{c|}{Non-thermal DM}
 \\
\hline
 {Parameter} & & {$\tau=10^2$s} & {$\tau=10^3$s} & {$\tau=10^4$s} 
\\
\hline
Primary & & & & \\
{$\Omega_b h^2$} 
& $0.02235 \pm 0.00013$
& $0.02240 \pm 0.00013$
& $0.02240 \pm 0.00012$
& $0.02243 \pm 0.00013$	
\\
{$\Omega_{c} h^2$} 
& $0.1172 \pm 0.0021 $
& $0.1199 \pm 0.0008 $ 
& $0.1201 \pm 0.0008$
& $0.1226 \pm 0.0008$
\\
{$100\theta$}
& $1.04126 \pm 0.00036 $
& $1.04098 \pm 0.00029 $
& $1.04103 \pm 0.00029$
& $1.04215 \pm 0.00069$
\\
{$\tau_{opt}$}
& $0.0784 \pm 0.0038$
& $0.0748 \pm 0.0027 $
& $0.0748 \pm 0.0027$
& $0.0731 \pm 0.0027$
\\
{$\omega$}
& $-1.034 \pm 0.022$
& $-1.036 \pm 0.023$
& $-1.037 \pm 0.024$
& $-1.043 \pm 0.024$
\\
{$\log_{10}\left(f\frac{m_{\chi'}}{m_{\chi}}\right)$}
& $ - $
& $<-0.726$
& $<0.861$
& $1.587 \pm 0.014$
\\
\hline
\hline
Derived & & & & \\
{$H_{0}$}
& $67.82 \pm 0.77 $
& $68.36 \pm 0.68$
& $68.41 \pm 0.69$
& $69.08 \pm 0.71$
\\
{$\Omega_{m}$}
& $0.3049 \pm 0.0063 $
& $0.3057 \pm 0.0066 $
& $0.3060 \pm 0.0064$
& $0.3057 \pm 0.0066$
\\
{$\Omega_{\Lambda}$}
& $0.6951 \pm 0.0063 $
& $0.6943 \pm 0.0066 $
& $0.6940 \pm 0.0064$
& $0.6943 \pm 0.0066$
\\
{$\sigma_8$}
& $0.833 \pm 0.009 $
& $0.839 \pm 0.008$
& $0.839 \pm 0.008$
& $0.842 \pm 0.008$
\\
{$S_8$}
& $0.840 \pm 0.010$
& $0.847 \pm 009$
& $0.847 \pm 009$
& $0.850 \pm 008$
\\
{$r_{drag}$}
& $148.6 \pm 0.1 $
& $147.2 \pm 0.2$
& $147.1 \pm 0.2$
& $146.4 \pm 0.2$
\\
\hline
{$\Delta N_{eff}$}
& $-0.136$
& $<4.69\times 10^{-6}$
& $<5.74\times 10^{-4} $
& $0.01 \pm 5\times 10^{-8}$
\\
\hline
\end{tabular} \label{tab:Tabel_results_1}}
}
\end{table*} 

\begin{figure*}[tbp]
    \centering
    \includegraphics[width=\textwidth]{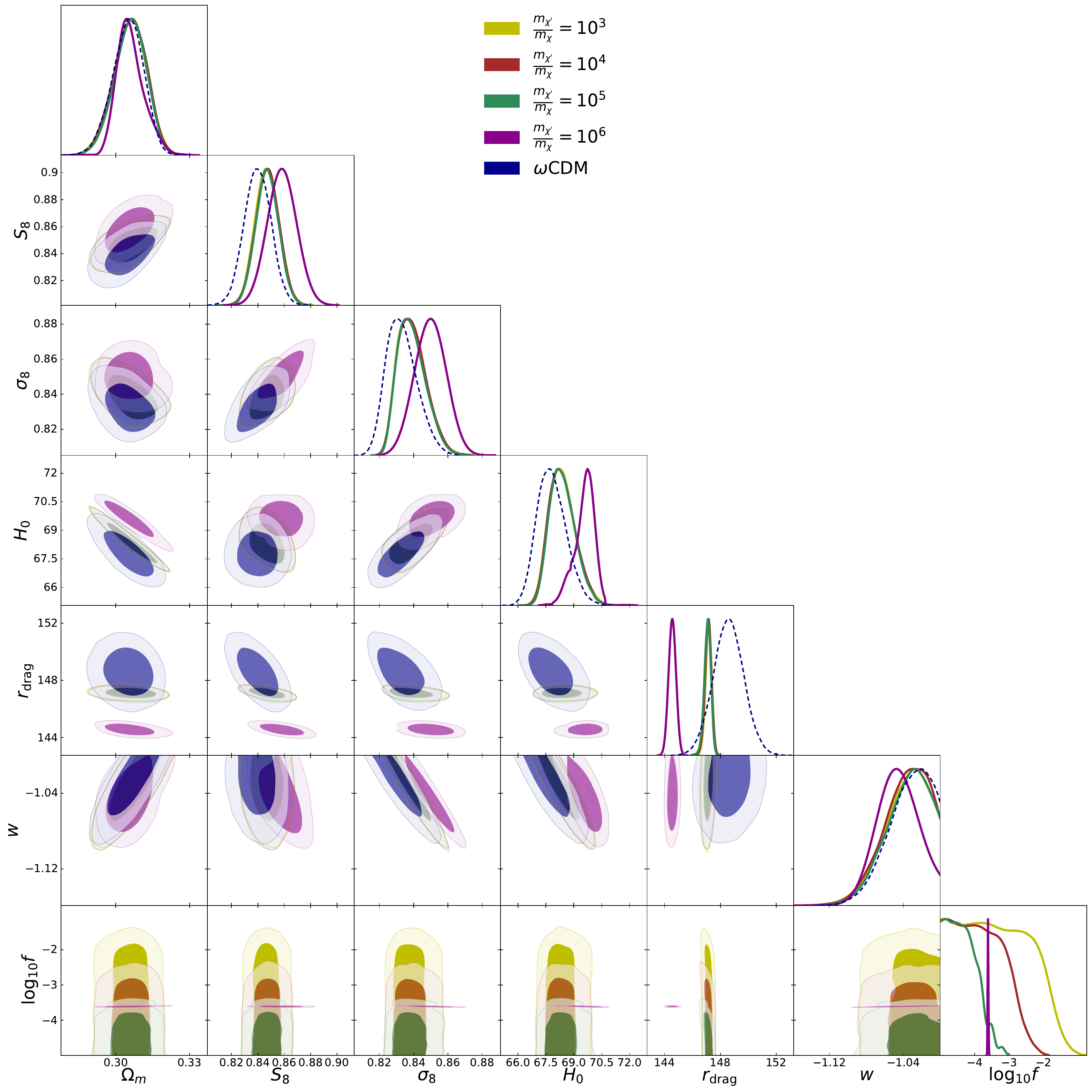}
    \caption{One-dimensional posterior distributions and two-dimensional joint contours for the parameter space $\{\Omega_bh^2, \Omega_ch^2, 100\theta, \tau, \omega, f\}$ for the lifetime of the dark-matter mother particle of $\tau=10^3$s and different values of $\frac{m_{\chi'}}{m_{\chi}}$, using the Planck 2018 + BAO + Pantheon dataset. Once again, we display the $\omega$CDM model for comparison purposes.}
    \label{fig:tri_tau3_f}
\end{figure*}

Motivated by the success in constraining the combination $f\frac{m_{\chi'}}{m_{\chi}}$, we decided to go further and split this combination as an attempt to constrain the fraction of hot dark matter particles produced due to the decay mechanism presented before, for fixed values of the mass ratio. The values for the mass ratio were chosen according to an upper limit of hot dark particles present at the matter-radiation equality. As you can see in \fig{fig:t_x_rho_ratio}, the higher the mass ratio, the higher the energy of the hot dark matter particles at the matter-radiation equality. If we have too many hot particles at this time, we risk destroying the success of the BBN and structure formation observables, whereas keeping fewer hot particles will maintain the expansion history unaltered. In this way, we chose to fix the mass ratio to the intermediate values of $\frac{m_{\chi'}}{m_{\chi}}=10^3, 10^4, 10^5$ and $10^6$. We proceeded similarly, adapting CAMB and CosmoMC codes for our model, fixing $\tau=10^2, 10^3, 10^4 s$ and $\frac{m_{\chi'}}{m_{\chi}}$ to the values previously discussed and varying all the other parameters. The results are displayed in \tab{tab:Tabel_results_2} and \fig{fig:fxH0_tau2}, \fig{fig:fxH0_tau3}, \fig{fig:fxH0_tau4}, and  \fig{fig:tri_tau3_f}.

In general, we obtain a very good agreement in $1\sigma$ with $\omega$CDM + $N_{eff}$ model for all values of $\tau$ and $\frac{m_{\chi'}}{m_{\chi}}$ considered, except the combination $\tau=10^3 s$ with $\frac{m_{\chi'}}{m_{\chi}}=10^6$, which present a very low value for $r_{drag}$ that is not in agreement with any other model, as you can observe in \fig{fig:tri_tau3_f}. We could expect that the small value for this early time parameter would be able to help us to solve the Hubble tension or, in the worst case, alleviate the $H_0$ tension but increase the $\sigma_8$ tension. However, neither of them happens. The reason for this is that higher values of $H_0$ present a very limited significance for the posterior probability distribution, in such a way that the data will always prefer smaller values for $H_0$. The most evident impact of changing $r_{drag}$ to smaller values, is the shift on the peak of $H_0$ posterior, which is moderately higher than the other models. This shift brings $H_0$ to $69.56\pm 0.42$, which means a reduction to $3.2\sigma$ tension concerning the local measurements.
Note that considering $f$ and the mass ratio $\frac{m_{\chi'}}{m_{\chi}}$ as independent parameters did not produce a satisfactory enhancement on $H_0$ to solve the current tension, but on the other hand, the decrease produced in $r_{drag}$ (for $\tau=10^3 s$ and $\frac{m_{\chi'}}{m_{\chi}}=10^6$) did not make the clustering parameter tension worse. Therefore, we could alleviate the $H_0$ tension without worsening the $\sigma_8$ tension. In summary, we have shown that a non-thermal dark matter plus a phanton-like cosmology can alleviate the $H_0$ trouble without creating another in different cosmological observables. if $H_0 > 72 km s^{-1}Mpc^{-1}$, then a new mechanism needs to be at play, other than non-thermal dark matter production and phantom-like cosmology.
\begin{table*}[tbp]
\centering
\caption{$68\%$ confidence limits for the cosmological parameters. The columns-block are divided according to the ratio, $\frac{m_{\chi^\prime}}{m_{\chi}}$, and the lifetime of the dark matter mother particle, $\tau_{life}$, values. The table shows the constraints on the parameters of the model using the extended data set, i.e. the joint Planck 2018 $+$ BAO $+$ Pantheon dataset. 
Note that the table is divided into two sections: the upper section shows the primary parameters, while the lower part shows the derived ones.}
\scalebox{0.65}{
{\begin{tabular}{|c|c|c|c|c|c|c|}
\hline
 &\multicolumn{3}{c|}{$\frac{m_{\chi'}}{m_{\chi}}=10^3$}
 &\multicolumn{3}{c|}{$\frac{m_{\chi'}}{m_{\chi}}=10^4$}
 \\
\hline
 {Parameter} & {$\tau=10^2$s} & {$\tau=10^3$s} & {$\tau=10^4$s} 
 &{$\tau=10^2$s} & {$\tau=10^3$s} & {$\tau=10^4$s}
 \\
\hline
Primary & & & & & &\\
{$\Omega_b h^2$} 
& $0.02240 \pm 0.00013$
& $0.02239 \pm 0.00013$
& $0.02240 \pm 0.00012$
& $0.02242 \pm 0.00013$	
& $0.02239 \pm 0.00013$
& $0.02240 \pm 0.00013$
\\
{$\Omega_{c} h^2$} 
& $0.1199 \pm 0.0008 $
& $0.1199 \pm 0.0008 $ 
& $0.1199 \pm 0.0008$
& $0.1213 \pm 0.0007$
& $0.1199 \pm 0.0008$
& $0.1200 \pm 0.0008$
\\
{$100\theta$}
& $1.04100 \pm 0.00029 $
& $1.04097 \pm 0.00029 $
& $1.04098 \pm 0.00029$
& $1.04149 \pm 0.00039$
& $1.04099 \pm 0.00029$
& $1.04100 \pm 0.00030$
\\
{$\tau_{opt}$}
& $0.0749 \pm 0.0027$
& $0.0748 \pm 0.0027 $
& $0.0748 \pm 0.0027$
& $0.0731 \pm 0.0027$
& $0.0748 \pm 0.0027$
& $0.0748 \pm 0.0027$
\\
{$\omega$}
& $-1.037 \pm 0.023$
& $-1.037 \pm 0.024$
& $-1.037 \pm 0.023$
& $-1.039 \pm 0.024$
& $-1.037 \pm 0.024$
& $-1.037 \pm 0.023$
\\
{$\log_{10}f$}
& $-2.56 \pm 1.07$
& $<-2.83$
& $-3.64 \pm 0.82$
& $-1.71 \pm 0.02$
& $<-3.5$
& $<-3.82$
\\
\hline
\hline
Derived & & & & & & \\
{$H_{0}$}
& $68.38 \pm 0.68$
& $68.39 \pm 0.69$
& $68.37 \pm 0.67$
& $68.73 \pm 0.70$
& $68.39 \pm 0.68$
& $68.40 \pm 0.68$
\\
{$\Omega_{m}$}
& $0.3059 \pm 0.0064$
& $0.3057 \pm 0.0066 $
& $0.3060 \pm 0.0064$
& $0.3057 \pm 0.0066$
& $0.3058 \pm 0.0065$
& $0.3058 \pm 0.0065$
\\
{$\Omega_{\Lambda}$}
& $0.6941 \pm 0.0064 $
& $0.6943 \pm 0.0066 $
& $0.6940 \pm 0.0064$
& $0.6943 \pm 0.0066$
& $0.6942 \pm 0.0066$
& $0.6942 \pm 0.0064$
\\
{$\sigma_8$}
& $0.839 \pm 0.008$
& $0.839 \pm 0.008$
& $0.839 \pm 0.008$
& $0.842 \pm 0.008$
& $0.839 \pm 0.008$
& $0.839 \pm 0.008$
\\
{$S_8$}
& $0.847 \pm 009$
& $0.847 \pm 009$
& $0.847 \pm 009$
& $0.850 \pm 008$
& $0.847 \pm 009$
& $0.847 \pm 009$
\\
{$r_{drag}$}
& $147.1 \pm 0.2$
& $147.2 \pm 0.2$
& $147.1 \pm 0.2$
& $146.4 \pm 0.2$
& $147.1 \pm 0.2$
& $147.1 \pm 0.2$
\\
\hline
\hline
\hline
 &\multicolumn{3}{c|}{$\frac{m_{\chi'}}{m_{\chi}}=10^5$}
 &\multicolumn{3}{c|}{$\frac{m_{\chi'}}{m_{\chi}}=10^6$}
 \\
\hline
 {Parameter} & {$\tau=10^2$s} & {$\tau=10^3$s} & {$\tau=10^4$s} 
 &{$\tau=10^2$s} & {$\tau=10^3$s} & {$\tau=10^4$s}
 \\
\hline
Primary & & & & & &\\
{$\Omega_b h^2$} 
& $0.02240 \pm 0.00012$
& $0.02240 \pm 0.00013$
& $0.02240 \pm 0.00012$
& $0.02240 \pm 0.00013$	
& $0.02246 \pm 0.00013$
& $0.02240 \pm 0.00013$
\\
{$\Omega_{c} h^2$} 
& $0.1206 \pm 0.0008 $
& $0.1200 \pm 0.0008 $ 
& $0.1206 \pm 0.0008$
& $0.1206 \pm 0.0008$
& $0.1250 \pm 0.0008$
& $0.1201 \pm 0.0008$
\\
{$100\theta$}
& $1.04122 \pm 0.00032 $
& $1.04103 \pm 0.00030 $
& $1.04122 \pm 0.00031$
& $1.04120  \pm 0.00031$
& $1.04389  \pm 0.00156$
& $1.04106  \pm 0.00031$
\\
{$\tau_{opt}$}
& $0.0740 \pm 0.0027$
& $0.0748 \pm 0.0027$
& $0.0739 \pm 0.0026$
& $0.0740 \pm 0.0027$
& $0.0689 \pm 0.0026$
& $0.0746 \pm 0.0027$
\\
{$\omega$}
& $-1.037 \pm 0.023$
& $-1.036 \pm 0.023$
& $-1.037 \pm 0.024$
& $-1.039  \pm 0.025$
& $-1.044  \pm 0.018$
& $>1.047$
\\
{$\log_{10}f$}
& $-3.02 \pm 0.31$
& $<-4.16$
& $-4.02 \pm 0.03$
& $-4.02 \pm 0.03$
& $-3.61 \pm 0.01$
& $<-5.46 $
\\
\hline
\hline
Derived & & & & & & \\
{$H_{0}$}
& $68.52 \pm 0.68$
& $68.40 \pm 0.67$
& $68.52 \pm 0.68$
& $68.57 \pm 0.71$
& $69.56 \pm 0.42$
& $68.44 \pm 0.69$
\\
{$\Omega_{m}$}
& $0.3061 \pm 0.0066$
& $0.3059 \pm 0.0063 $
& $0.3061 \pm 0.0065$
& $0.3056 \pm 0.0068$
& $0.3061 \pm 0.0043$
& $0.3057 \pm 0.0065$
\\
{$\Omega_{\Lambda}$}
& $0.6939 \pm 0.0066 $
& $0.6941 \pm 0.0063 $
& $0.6939 \pm 0.0065$
& $0.6944 \pm 0.0068$
& $0.6939 \pm 0.0043$
& $0.6943 \pm 0.0065$
\\
{$\sigma_8$}
& $0.840 \pm 0.008$
& $0.839 \pm 0.008$
& $0.840 \pm 0.008$
& $0.841 \pm 0.008$
& $0.850 \pm 0.007$
& $0.839 \pm 0.008$
\\
{$S_8$}
& $0.849 \pm 009$
& $0.847 \pm 008$
& $0.849 \pm 008$
& $0.848 \pm 009$
& $0.858 \pm 008$
& $0.847 \pm 009$
\\
{$r_{drag}$}
& $146.8 \pm 0.2$
& $147.1 \pm 0.2$
& $146.8 \pm 0.2$
& $146.8 \pm 0.2$
& $144.6 \pm 0.2$
& $147.0 \pm 0.3$
\\
\hline
%
\end{tabular} \label{tab:Tabel_results_2}}
}
\end{table*} 
%
%
%
%
%
%

\section{\label{dc} Conclusions}

In this work, we evaluate the cosmological consequences of the decay of a heavy particle into dark matter and neutrinos ($\chi^\prime \rightarrow \chi + \nu$). As $m_{\chi^\prime} \gg m_{\chi}$, the dark matter particles are initially relativistic and contribute to the radiation energy density. This results in $\Delta N_{eff}\neq 0$. Although, this mechanism is capable of increasing $N_{eff}$, it does not suffice to raise $H_0$ to values consistent with the local measurements. In order to properly address this issue, we performed a Monte Carlo simulation using the CAMB code to assess whether a non-thermal dark matter production aided by a phantom-like fluid as a background foots the bill without altering much other cosmological parameters. In our analysis, we used the CMB, BAO and data from type Ia supernovae datasets. We found that such a setup can yield $H_0=69.08\pm0.71$ Km s$^{-1}$ Mpc$^{-1}$ for $\tau=10^4$, which alleviates the tension between early and late measurements of the Hubble constant to $3.2\sigma$. We highlight that the overall dark matter abundance cannot stem from this mechanism. According to data from Sloan Digital Sky Survey at most 6\% of the overall dark matter abundance can be relativistic when structures start to be formed. With that in mind, we considered the fraction of relativistically produced dark matter as a free parameter, we set a limit on it. We limited this fraction to be at most 1\% in many cases. Thus, our bound is more restrictive than those stemming from structure formation. 

We emphasize that the mechanism employed here introduces a minor correction to the $\Lambda$CDM model. In other words, the central values of the cosmological parameters do not deviate so much from the standard case. This is important because we do not expect any substantial change in the well-established and data-fitted standard cosmological scenarios. In summary, we concluded that the Hubble constant can serve as an early universe probe for dark sectors. 

\acknowledgments
The authors thank Jacinto Paulo for discussions. ASJ acknowledges support from Coordenaç\~ao de Aperfeiçoamento de Pessoal de N\'ivel Superior (CAPES) under Grant No. 88887.497142/2020-00. DRS thanks for the support from CNPQ under grant 303699/2023-0. SSC is supported by the Istituto Nazionale di Fisica Nucleare (INFN), sezione di Pisa, iniziativa specifica TASP. NPN acknowledges the support of CNPq of Brazil under grant PQ-IB 310121/2021-3.  FSQ is supported by Simons Foundation (Award Number:1023171-RC), FAPESP Grant 2018/25225-9, 2021/01089-1, 2023/01197-4, ICTP-SAIFR FAPESP Grants 2021/14335-0, CNPq Grants 307130/2021-5, and ANID-Millennium Science Initiative Program ICN2019\textunderscore044.

\nocite{*}

\appendix

\section{Entropy injection bounds}
\label{sec:entropy_injection}

In this appendix, we show that such non-thermal production of dark matter does add much entropy to the universe. We are considering a non-thermal decay of $\chi^\prime$. Therefore, this decay will introduce entropy into the primordial plasma. For this mechanism to be useful, it must not introduce an appreciable amount of entropy. In this section, we demonstrate the correlation between the relative entropy added and $\Delta N_{\mathrm{eff}}$. We also show that the mechanism used here does not significantly increase entropy and remains within this bound.

Let's begin by calculating the total entropy density, which is the sum of the fermionic and bosonic contributions \cite{dodelson2020}, i.e.,
\begin{equation}
    s_{\text{rad}} = \sum_b \frac{2\pi^2 g_b}{45}T^3_b + \frac{7}{8} \sum_f \frac{2\pi^2 g_f}{45}T^3_f.
\end{equation}

It is usual to define
\begin{equation}
    g_{*s}(T) = \sum_b g_b \left( \frac{T_b}{T} \right)^3 + \frac{7}{8} \sum_f g_f\left( \frac{T_f}{T} \right)^3.
    \label{eq:g*s}
\end{equation}
Here, $T$ is the photons' temperature.

The above definition rewrites the total radiation entropy density as follows:
\begin{equation}
    s_{rad} = \frac{2\pi^2 g_{*s}}{45}T^3.
    \label{eq:s_rad}
\end{equation}

Eq. (\ref{eq:s_rad}) tells us that the entropy ratio between two moments $t_i$ and $t_f$ is given by
\begin{equation}
    \frac{S_f}{S_i} = \frac{g_{*s}^f T_f^3 a_f^3}{g_{*s}^i T_i^3 a_i^3}\cdot
    \label{eq:S_ratio_temperature}
\end{equation}

In the above identity, we have both time and temperature as dynamic parameters. It is advisable to choose only one of them. To do that, we can utilize the fact that during the radiation-dominated phase, the Hubble parameter is given by $H = 1/(2t)$, and Friedman's equation yields $H^2 = 8\pi\rho_r/3$ \cite{dodelson2020}. Both of these equations can be used to express the radiation energy density as follows:
\begin{equation}
    \rho_r = \frac{3}{32\pi t^2}\cdot
\end{equation}

The total radiation energy is  $\rho_r = \pi^2 g_{*} T^4/30$ \cite{dodelson2020}, where the factor $g_{*}$ is defined by
\begin{equation}
    g_{*}(T) \equiv \sum_b g_b \left( \frac{T_b}{T} \right)^4 + \frac{7}{8} \sum_f g_f \left( \frac{T_f}{T} \right)^4.
\end{equation}

We have two expressions for radiation energy density, one that is an explicit function of time and another that is an explicit function of temperature. Equating these formulas provides a direct connection between temperature and time at the radiation-dominated phase:
\begin{equation}
    T = \left( \frac{45}{16\pi^3g_{*} t^2} \right)^{1/4}.
    \label{eq:T}
\end{equation}

With this result, we can express the ratio $T_f / T_i$ in Eq. (\ref{eq:S_ratio_temperature}) as follows:
\begin{equation}
    \frac{T_i}{T_f} = \left( \frac{g^f_{*}}{g_{*}^i} \right)^{1/4}\sqrt{\frac{t_f}{t_i}}\cdot
\end{equation}

In Eq. (\ref{eq:S_ratio_temperature}) the ratio $a_f / a_i$ can also be expressed as a function of the time ratio:
\begin{equation}
    \frac{a_f}{a_i} = \left( \frac{t_f}{t_i} \right)^{1/2}.
\end{equation}

Applying the temperature ratio and the scale factor ratio to the entropy ratio (Eq. \ref{eq:S_ratio_temperature}) yields:
\begin{equation}
    \frac{S_f}{S_i} = \frac{g_{*s}^f}{g_{*s}^i} \left( \frac{g_{*}^i}{g_{*}^f} \right)^{3/4}\cdot
    \label{eq:entropy_ratio}
\end{equation}

The immediate consequence of this result is that in standard cosmology, between BBN and the CMB, $g_{*s}^i = g_{*s}^f$ and $g_{*}^i = g_{*s}^i$, which implies that between these two events, there is no relevant addition to the entropy radiation, i.e., $S_f = S_i$ \cite{Kaplinghat2001, ref:Alcaniz2021, Feng:2003uy, kolb2018early}.

However, we are not considering the standard cosmological scenario. The decay of $\chi^\prime$ adds an effective number of neutrinos, increasing the entropy of the fundamental plasma. Eq. (\ref{eq:entropy_ratio}) informs that the relative entropy variation is given by:
\begin{equation}
    \frac{S_f - S_i}{S_i} = \frac{\Delta S}{S_i} = \frac{g_{*s}^f}{g_{*s}^i} \left( \frac{g_{*}^i}{g_{*}^f} \right)^{3/4} - 1.
    \label{eq:entropy_variation}
\end{equation}

Here, the gs factors are defined by
\begin{subequations}
    \begin{equation}
        \begin{split}
            g_{*}^{i} &= g_ \gamma + \frac{7}{8} N_\nu \times 2 \times g_\nu  \left( \frac{T_\nu}{T_\gamma} \right)^4 = 2 + \frac{21}{4}  \left( \frac{4}{11} \right)^{4/3} \\
            & = 3.36264,
        \end{split}
    \end{equation}
    \begin{equation}
        \begin{split}
            g_{*s}^{i} & = g_ \gamma + \frac{7}{8} \times N_\nu \times 2 \times g_\nu \left( \frac{T_\nu}{T_\gamma} \right)^3 = 2 + \frac{21}{4} \left( \frac{4}{11} \right)\\
            & = 3.90909,
        \end{split}
    \end{equation}
    \begin{equation}
        \begin{split}
            g_{*}^{f} & = g_ \gamma + \frac{7}{8} (3 + \Delta N_{\text{eff}}) \times 2 \times g_\nu  \left( \frac{T_\nu}{T_\gamma} \right)^4 \\
            & = 2 + \frac{7}{4} (3 + \Delta N_{\text{eff}})  \left( \frac{4}{11} \right)^{4/3},
        \end{split}
    \end{equation}
    \begin{equation}
        \begin{split}
            g_{*s}^{f} & = g_ \gamma + \frac{7}{8} \times (3 + \Delta N_{\text{eff}}) \times 2 \times g_\nu \left( \frac{T_\nu}{T_\gamma} \right)^3\\
            & =  2 + \frac{7}{4} \times (3 + \Delta N_{\text{eff}}) \left( \frac{4}{11} \right)\cdot
        \end{split}
    \end{equation}
\end{subequations}

In the factors mentioned above, we considered that initially only background neutrinos and photons are relativistic. After the decay of $\chi^\prime$, these gs factors changed. This is the reason the contribution to $\Delta N_{\mathrm{eff}}$ arises. It should also be clear that we are assuming that the background neutrinos have already decoupled from photons. Therefore, we used the ratio between the neutrino and photon temperatures, $T_\nu / T_\gamma = (4 / 11)^{1/3}$ \cite{dodelson2020}.
\begin{figure}[tbp]
\centering
\includegraphics[width=0.5\textwidth]{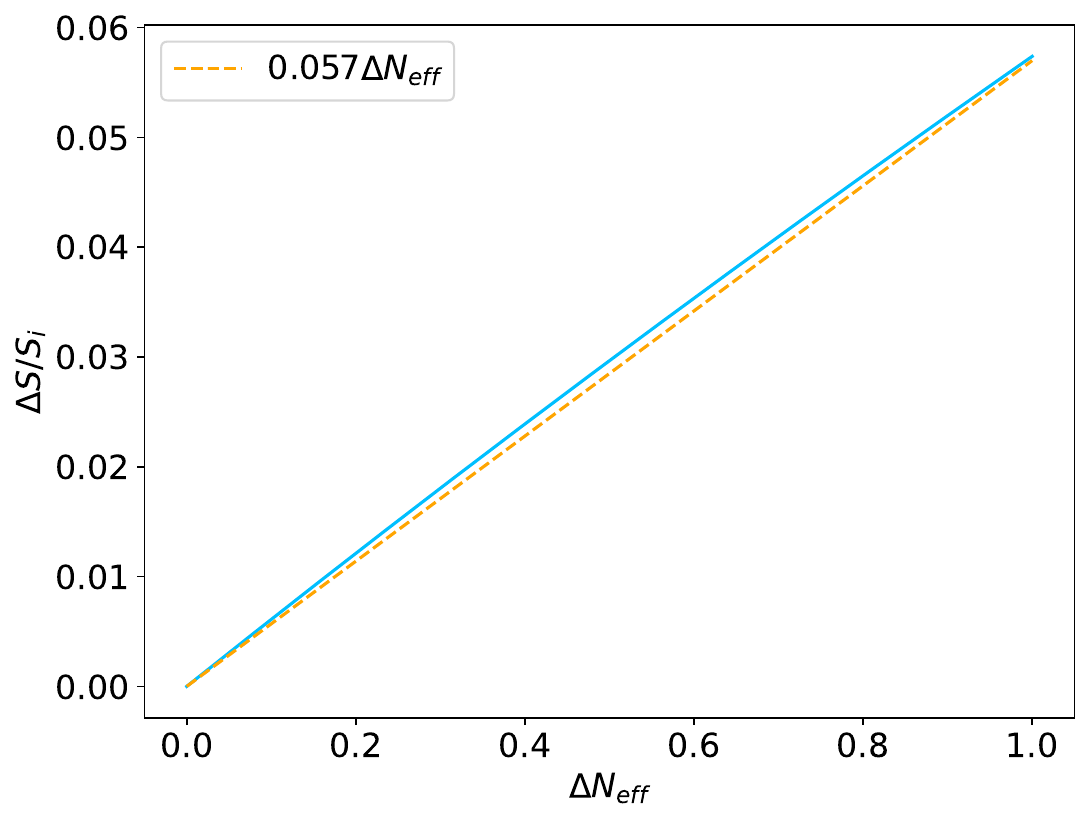}
\caption{Relative entropy variation as a function of $\Delta N_\text{eff}$. The blue solid line is obtained by Eq. \ref{eq:entropy_variation}, while the orange dotted line is a linear approximation.}
\label{fig:entropy_variation}
\end{figure}

It is more elucidative to deal with the $\Delta S / S_i$ ratio numerically. In Fig. \ref{fig:entropy_variation}, we demonstrate how the relative entropy variation changes with $\Delta N_{\text{eff}}$. It is important to note that in the range $0 \leq \Delta N_{\text{eff}} \leq 1$, the approximation
\begin{equation}
    \frac{\Delta S}{S_i} = 0.057 \Delta N_{\text{eff}}
    \label{eq:entropy_variation_and_DeltaNeff}
\end{equation}
is quite accurate. Furthermore, observe that for the upper limit, $\Delta N_{\text{eff}} = 1$, the relative entropy variation is less than $6\%$. This is an excellent indication that the formalism used here does not add too much entropy in agreement with BBN and CMB observations.


%

\end{document}